\documentclass[final,1p,times]{elsarticle}

\usepackage{amssymb}
\usepackage{threeparttable}
\newcommand{\ket}[1]{\mbox{$ | #1 \rangle $}}

\begin{document}

\begin{frontmatter}

\title{Semiquantum secret sharing using two-particle entangled state}


\author{Jian Wang\fnref{fn1}}
\author{Sheng Zhang}
\author{Quan Zhang}
\author{Chao-Jing Tang}

\address{School of Electronic Science and Engineering, National University of Defense Technology, Changsha 410073, China}
\fntext[fn1]{Tel. No.: 86 0731 84575706, E-mail address: jwang@nudt.edu.cn}

\begin{abstract}
Recently, Boyer et al. presented a novel semiquantum key distribution protocol
[M. Boyer, D. Kenigsberg, and T. Mor, Phys. Rev. Lett. 99, 140501 (2007)], in which
quantum Alice shares a secret key with classical Bob. Li et al. proposed
two semiquantum secret sharing protocols [Q. Li, W. H. Chan, and D. Y. Long, Phys. Rev. A 82,
022303 (2010)] by using maximally entangled Greenberger-Horne-Zeilinger states.
In this paper, we present a semiquantum secret sharing protocol by using two-particle entangled states
in which quantum Alice shares a secret key with two classical parties, Bob and Charlie.
Classical Bob and Charlie are restricted to performing measurement in the computational basis,
preparing a particle in the computational basis, or reflecting the particles. None of them
can acquire the secret unless they collaborate. We also show the protocol is secure against eavesdropping.
\end{abstract}


\end{frontmatter}


\section{Introduction}
The basic idea of secret sharing in the simplest case is that the sender,
Alice, splits the secret message into two shares and distributes
them to two receivers, Bob and Charlie, respectively, such that only
the two receivers collaborate can they reconstruct the secret
message. In a more general setting, an $(m, n)$ threshold scheme,
the secret message is split into $n$ shares, such that any $m$ of
those shares can be used to reconstruct it. Quantum secret sharing (QSS) is the
generalization of classical secret sharing and can share both
classical and quantum message. QSS is likely to play a key role in
protecting secret quantum information, e.g., in secure operations of
distributed quantum computation, sharing difficult-to-construct
ancillary states and joint sharing of quantum money, etc.

Many researches have been carried out in both theoretical and
experimental aspects after the pioneering QSS scheme proposed by
Hillery, Buz\v{e}k and Berthiaume\cite{HBB99} in 1999. HBB99 scheme is based on a three-particle
Greenberger-Horne-Zeilinger (GHZ) state. Karlsson, Koashi and Imoto
\cite{KKI99} proposed a QSS scheme using two-particle Bell states.
Guo-Ping Guo and Guang-Can Guo\cite{GG03} presented a QSS scheme
where only product states are employed. Li Xiao et al.\cite{XLDP04} generalized the HBB99 scheme
into arbitrary multiparties and improved the efficiency of the QSS
scheme by using two techniques from quantum key distribution. Zhan-jun
Zhang et al.\cite{ZLM05} proposed an ($n, n$) threshold scheme of multiparty QSS of classical messages using only
single photons. Fu-guo Deng et al.\cite{DLZZ05} improved the security of multiparty QSS against
Trojan horse attack with two single-photon measurements and four unitary operations. Zhan-jun Zhang and Zhong-xiao Man put forward a
multiparty QSS protocol by using swapping quantum entanglement of Bell states\cite{ZM05} and a multiparty QSS protocol of secure direct
communication based on the two-step QSDC protocol\cite{Z-PLA05}.

Recently, Boyer et al. presented a novel semiquantum key distribution protocol\cite{BKM07} (hereafter we call it BKM07 protocol),
in which one party (Alice) is quantum and the other (Bob) is classical,
and proved that the protocol is completely robust against an eavesdropping attempt.
In their protocol, quantum Alice prepares a random qubit in the computational ($Z$) basis \{\ket{0},\ket{1}\} or
Hadamard ($X$) basis \{$\ket{+}=\frac{1}{\sqrt{2}}(\ket{0}+\ket{1})$, \ket{-}=$\frac{1}{\sqrt{2}}(\ket{0}-\ket{1})$\}
and sends it to Bob. They call the computational basis classical
and use the classical notion \{0,1\} to describe the two quantum states
\{\ket{0},\ket{1}\}. Classical Bob is restricted to performing some classical operations, such as
measuring the transmission qubits in the classical \{0,1\} basis, preparing a qubit in the classical basis and sending it,
reflecting the particle directly. Boyer et al. showed a different semiquantum key distribution protocol\cite{BGKM09}
based on randomization. Different from BKM07 protocol, classical Bob can reorder the particles besides measuring
and preparing a qubit in the classical basis. In this protocol, Bob reorders randomly the reflected
qubits in order to avoid Eve's acquiring Bob's operation information on each receiving qubit.
Furthermore, they proved the robustness of the protocol in much more general scenario.
Zou et al. presented five different semiquantum key distribution protocols\cite{ZOU09} in which Alice sends three quantum states, two
quantum states and one quantum state, respectively. Li et al. proposed two semiquantum secret sharing protocols (SQSS)\cite{LI10}
(hereafter we call it LCL10 protocol) by using maximally entangled Greenberger-Horne-Zeilinger states in which quantum Alice shares a secret
with two classical parties, Bob and Charlie. In LCL10 protocol, quantum Alice prepares a batch of three-particle entangled state,
each of which is in the state $\ket{\psi}=\frac{1}{\sqrt{2}}(\ket{0}\frac{\ket{00}+\ket{11}}{\sqrt{2}}+\ket{1}\frac{\ket{01}+\ket{10}}{\sqrt{2}})$,
and sends the second and the third particles of each entangled state to Bob and Charlie. By utilizing the method of randomization or measure-resend, classical Bob and Charlie can share a secret with quantum Alice. Neither Bob nor Charlie can reconstruct Alice's secret
unless they collaborate with each other.

In this paper, we present a SQSS protocol by using two-particle entangled state.
We follow the descriptions about classical in Ref.\cite{BKM07}.
Quantum Alice can prepare two-particle entangled states and measure the particles
in the computational basis, Hadamard basis, or Bell basis. Classical Bob and Charlie are restricted to
measuring the particle in the computational basis, preparing a qubit in the computational basis, sending or reflecting the particles.
We show that Eve's eavesdropping would inevitably disturb
the transmission quantum states and the communication parties can detect
Eve's attack.

%

\section{The description of the SQSS protocol}
Suppose the sender, Alice, wants to share a secret key with two receivers, Bob and
Charlie, so that none of them can recover the secret
message on his own. The protocol is detailed as follows:

(1) Alice prepares $N$ two-particle entangled states, each of which is
in the state
\begin{eqnarray}
\ket{\Psi}&=&\frac{1}{\sqrt{2}}(\ket{+0}+\ket{-1})_{BC},
\end{eqnarray}
where $\ket{+}=\frac{1}{\sqrt{2}}(\ket{0}+\ket{1})$, \ket{-}=$\frac{1}{\sqrt{2}}(\ket{0}-\ket{1})$.
We denote the ordered $N$ two-particle entangled states by
\{[P$_1(B)$,P$_1(C)$], [P$_2(B)$,P$_2(C)$],
$\cdots$, [P$_N(B)$,P$_N(C)$]\}, where the subscript
indicates the order of each state in the sequence, and $B$, $C$
represent the two particles of each state. Alice takes
particle B from each state to form an ordered partner particle
sequence, [P$_1(B)$, P$_2(B)$,$\cdots$, P$_N(B)$], called $B$
sequence. The remaining partner particles comprise $C$ sequence,
[P$_1(C)$, P$_2(C)$,$\cdots$, P$_N(C)$]. Alice sends $B$ sequence and $C$
sequence to Bob and Charlie, respectively.

(2) When each particle arrives, Bob chooses randomly either to measure it
in the computational basis and resend it in the same state he found (we refer to this action as MEAS-RESEND),
or to reflect it to Alice directly (we refer to this action as REFLECT).
Similarly, Charlie selects randomly either to MEAS-RESEND
it or to REFLECT it directly the time each particle arrives.

(3) Alice stores the received particles in quantum memory and
informs Bob and Charlie that she has received the $B$ and $C$ sequence particles.
Bob and Charlie then publish which particles they chose to MEAS-RESEND and which ones they chose to REFLECT.

(4) Alice performs one of the four operations on each received particle
according to Bob's and Charlie's choices, as illustrated in Table 1.

\begin{table}[h]
\caption{The communication parties' operations on the particles}\label{Tab:one}
  \centering
    \begin{tabular}[b]{|c|c|c|c|} \hline
       Case & Bob & Charlie & Alice \\ \hline
       \ (i) & MEAS-RESEND & MEAS-RESEND & OPERATION1\tnote{1} \\ \hline
       \ (ii) & MEAS-RESEND & REFLECT & OPERATION2\tnote{2} \\ \hline
       \ (iii) & REFLECT & MEAS-RESEND & OPERATION3\tnote{3} \\ \hline
       \ (iv) & REFLECT & REFLECT & OPERATION4\tnote{4} \\ \hline
    \end{tabular}
    \begin{tablenotes}
     \item [1] OPERATION1: To measure particle B and C in $Z$ basis.
     \item [2] OPERATION2: To measure particle B in $Z$ basis and particle C in $X$ basis.
     \item [3] OPERATION3: To measure particle C in $Z$ basis and particle B in $X$ basis.
     \item [4] OPERATION4: To perform Hadamard transformation on particle B and then measure particle B and C in Bell basis.
    \end{tablenotes}
\end{table}

(i) If both Bob and Charlie select to MEAS-RESEND, Alice implements OPERATION1 to obtain the sifted secret message.
In this case, Bob and Charlie can share a secret key with Alice according to their measurement results,
since Alice can obtain both the parties' measurement results by implementing OPERATION1, as illustrated in Table 2.

\begin{table}[h]
\caption{The communication parties' measurement results and the shared secret key}\label{Tab:two}
  \centering
    \begin{tabular}[b]{|c|c|c|c|} \hline
       Bob's result & Charlie's result & Alice's results & Secret \\ \hline
       \ \ket{0} & \ket{0} & \ket{00} & 0 \\ \hline
       \ \ket{0} & \ket{1} & \ket{01} & 1 \\ \hline
       \ \ket{1} & \ket{0} & \ket{10} & 1 \\ \hline
       \ \ket{1} & \ket{1} & \ket{11} & 0 \\ \hline
    \end{tabular}
\end{table}

(ii) If Bob chooses to MEAS-RESEND and Charlie chooses to REFLECT, Alice carries out OPERATION2.
Since \ket{\Psi} can be rewritten as
\begin{eqnarray}
\ket{\Psi}&=&\frac{1}{\sqrt{2}}(\ket{0+}+\ket{1-})_{BC},
\end{eqnarray}
it changes to $\ket{0+}_{BC}$ or $\ket{1-}_{BC}$ after Bob and Charlie's operation.
Alice measures particle B in $Z$ basis and particle C in $X$ basis, and obtains $\ket{0+}_{BC}$ or $\ket{1-}_{BC}$.
She can check eavesdropping in the transmission line according to her measurement results.
That is to say, she checks whether the state of particle B and C is $\ket{0+}_{BC}$ or $\ket{1-}_{BC}$.

(iii) Suppose Bob selects to REFLECT and Charlie chooses to MEAS-RESEND.
After Bob and Charlie's operation, \ket{\Psi} collapses to $\ket{+0}_{BC}$ or $\ket{-1}_{BC}$.
Alice implements OPERATION3 and obtains $\ket{+0}_{BC}$ or $\ket{-1}_{BC}$.
In this case, she can check whether the state of particle B and C is right
according to her measurement results.

(iv) If both Bob and Charlie select to REFLECT, Alice performs OPERATION4.
After she performing Hadamard transformation on particle B, \ket{\Psi} is changed to
\begin{eqnarray}
\ket{\phi^+}&=&\frac{1}{\sqrt{2}}(\ket{00}+\ket{11})_{BC}.
\end{eqnarray}
Alice can then check whether the initial two-particle entangled state is destroyed.

(5) Alice checks the error rate in cases (ii), (iii), (iv). If the error rate is higher
than the threshold they preset, they abort the protocol.

(6) Alice checks the error rate in cases (i). She chooses randomly a sufficiently large subset
from the measurement results in case (i) and announces which are the chosen particles.
Bob and Charlie then publish their measurement results. Since Alice can obtain their
measurement results by implementing OPERAION1, she can check the error rate according to the information
announced by them. If it is below the threshold they preset, Bob and Charlie can then obtain the final shared secret key
which can only be reconstructed when they collaborate with each other.

Actually, the protocol can also be realized by using some other two-particle entangled states, such as
\begin{eqnarray}
\ket{\Phi}&=&-\frac{1}{\sqrt{2}}(\ket{+1}-\ket{-0})_{BC}.
\end{eqnarray}

\section{The security for the Protocol}
So far we have presented the SQSS protocol. We then discuss the security for the present protocol.
The key of the security of the protocol is to keep an eavesdropper, Eve or one dishonest party from knowing
which particles are MEAS-RESEND particles and which are REFLECT ones. If Eve or one dishonest party
cannot distinguish MEAS-RESEND and REFLECT particles before step (4), Alice's operation (OPERATION2 and OPERATION3)
is equal to performing random $Z$ basis measurement and $X$ basis measurement on particle B and C,
which can ensure the security of the protocol.
We show that if Eve or one dishonest party can acquire nonzero information about the secret message, the communication parties
can find errors by eavesdropping check with nonzero probability.

Suppose Bob is dishonest and he has managed to get Charlie's particles as
well as his own. We call dishonest Bob, Bob*. In this attack, we suppose Bob* has quantum capabilities.
He intercepts the particles in $C$ sequence and measures particles
B and C in $Z$-basis, $X$-basis, or Bell basis. He then
resends $C$ sequence to Charlie after measurements.

\subsection{Bob* measures particle B and C in $Z$ basis}
When Bob* performs $Z$ basis measurement on particle B and C, the state of the whole system
collapses to $\ket{00}_{BC}$, $\ket{01}_{BC}$, $\ket{10}_{BC}$, or $\ket{11}_{BC}$, each with probability 1/4.
We then analyze the error rate introduced by Bob* in the four cases, respectively.

(1) In the case of both Bob* and Charlie choosing to MEAS-RESEND, Bob* can obtain Charlie's results
and his eavesdropping will not introduce any error at step (6).

(2) In case (ii), Bob* selects to MEAS-RESEND and Charlie selects to REFLECT.
Suppose \ket{\Psi} collapses to $\ket{00}_{BC}$ after Bob*'s eavesdropping attack. According to the protocol, Alice
measures particle B in $Z$ basis and particle C in $X$ basis.
After Alice's measurements, $\ket{00}_{BC}$ changes to
$\ket{0+}_{BC}$ or $\ket{0-}_{BC}$ each with probability 1/2.
As $\ket{\Psi}=\frac{1}{\sqrt{2}}(\ket{0+}+\ket{1-})_{BC}$, it is impossible for her to obtain
$\ket{0-}_{BC}$ and the error rate introduced by Bob* will reach 50\%.
Similarly, if \ket{\Psi} collapses to $\ket{01}_{BC}$, $\ket{10}_{BC}$, or $\ket{11}_{BC}$,
Bob*'s eavesdropping will also introduce an error rate of 50\%.

(3) In case (iii), Bob* selects to REFLECT and Charlie selects to MEAS-RESEND.
For example, \ket{\Psi} collapses to $\ket{00}_{BC}$ after Bob*'s eavesdropping attack.
Alice measures particle B in $X$ basis and particle C in $Z$ basis, and
obtains $\ket{+0}_{BC}$ or $\ket{-0}_{BC}$.
Since $\ket{\Psi}=\frac{1}{\sqrt{2}}(\ket{+0}+\ket{-1})_{BC}$, it is impossible for her to obtain $\ket{-0}_{BC}$.
Thus the error rate introduced by Bob* will reach 50\% in this case.

(4) In case (iv), both Bob and Charlie selects to REFLECT.
Alice first performs Hadamard transformation on particle B
and then measures particle B and C in Bell basis.
Suppose \ket{\Psi} collapses to $\ket{00}_{BC}$ after Bob*'s eavesdropping attack.
Alice performs Hadamard transformation on particle B and $\ket{00}_{BC}$ changes to
$\ket{+0}_{BC}$.
$\ket{+0}_{BC}$ can be rewritten as
\begin{eqnarray}
\ket{+0}_{BC}=\frac{1}{\sqrt{2}}(\ket{00}+\ket{10})_{BC}=\frac{1}{2}(\ket{\phi^+}+\ket{\phi^-}+\ket{\psi^+}-\ket{\psi^-})_{BC},
\end{eqnarray}
where
\begin{eqnarray}
\ket{\phi^-}=\frac{1}{\sqrt{2}}(\ket{00}-\ket{11}),\\
\ket{\psi^+}=\frac{1}{\sqrt{2}}(\ket{01}+\ket{10}),\\
\ket{\psi^-}=\frac{1}{\sqrt{2}}(\ket{01}-\ket{10}).
\end{eqnarray}
As none but \ket{\phi^+} is the right state, the error rate introduced by Bob* in this case is 3/4.

As discussed above, in the four cases, the average error rate introduce by Bob* is $\frac{1}{4}*(0+\frac{1}{2}+\frac{1}{2}+\frac{3}{4})=\frac{7}{16}=43.75\%$.

\subsection{Bob* measures particle B in $X$ basis and particle C in $Z$ basis}
Suppose Bob* measures particle B in $X$ basis and particle C in $Z$ basis, and
resends particle C to Charlie. After Bob*'s attack, the initial state collapses to
$\ket{+0}_{BC}$ or $\ket{-0}_{BC}$, each with probability 1/2.

(1) In case (i), both Bob* and Charlie measure their corresponding particle in $Z$ basis
and resend their particles to Alice directly. After Bob* and Charlie performing their operations,
the state of particle B and C collapses to $\ket{00}_{BC}$, $\ket{01}_{BC}$, $\ket{10}_{BC}$, or $\ket{11}_{BC}$,
each with probability 1/4. Bob* can then achieve Charlie's secret message and his eavesdropping
will not be detected at step (6), since Alice just performs $Z$ basis measurement on each arrived particle.

(2) In case (ii), Alice performs $Z$ basis measurement on particle B and $X$ basis measurement on particle C.
For example, if \ket{\Psi} collapses to $\ket{+0}_{BC}$ after Bob*'s eavesdropping, Alice obtains
$\ket{0+}_{BC}$, $\ket{0-}_{BC}$, $\ket{1+}_{BC}$, or $\ket{1-}_{BC}$, each with probability 1/4, after her measurements.
If there is no eavesdropping, Alice can only obtain $\ket{0+}_{BC}$ or $\ket{1-}_{BC}$.
Thus Bob*'s eavesdropping will introduce an error rate of 50\%.

(3) In case (iii), Alice measures particle B in $X$ basis and particle C in $Z$ basis.
After Bob*'s eavesdropping, \ket{\Psi} collapses to $\ket{+0}_{BC}$ or $\ket{-0}_{BC}$.
In this case, Bob*'s eavesdropping will not introduce any error.

(4) In case (iv), Alice performs Hadamard transformation on particle B and $\ket{+0}_{BC}$ ($\ket{-0}_{BC}$) becomes
$\ket{00}_{BC}$ ($\ket{10}_{BC}$). Alice then measures particle B and C in Bell basis.
As
$\ket{00}_{BC}=\frac{1}{\sqrt{2}}(\ket{\phi^+}+\ket{\phi^-})_{BC}$ and
$\ket{10}_{BC}=\frac{1}{\sqrt{2}}(\ket{\psi^+}-\ket{\psi^-})_{BC}$,
the error rate introduced by Bob* is 3/4.

As a result, the average error rate introduced by Bob* in the four cases is
$\frac{1}{4}*(0+\frac{1}{2}+0+\frac{3}{4})=\frac{5}{16}=31.25\%$.

In the case of Bob* performing $Z$ basis measurement on particle B and $X$ basis measurement on particle C, or measuring
particle B and C in $X$ basis, he cannot obtain Charlie's secret message and his eavesdropping will inevitably be
detected by Alice with a certain probability.

\subsection{Bob* measures particle B and C in Bell basis}
Suppose Bob* performs Bell basis measurement on particle B and C, and resends particle C to Charlie after measurement.
Owing to Bob*'s eavesdropping, \ket{\Psi} collapses to $\ket{\phi^-}_{BC}$ or $\ket{\psi^+}_{BC}$, each with probability 1/2.

(1) In case (i), both Bob* and Charlie selects to MEAS-RESEND. As
$\ket{\phi^-}_{BC}=\frac{1}{\sqrt{2}}(\ket{00}-\ket{11})_{BC}$ and
$\ket{\psi^+}_{BC}=\frac{1}{\sqrt{2}}(\ket{01}+\ket{10})_{BC}$, Bob* can acquire Charlie's secret message
according to his measurement result and his eavesdropping will not introduce any error.
For example, if \ket{\Psi} collapses to $\ket{\phi^-}_{BC}$ and Bob*'s measurement result is \ket{0}, he can infer that
Charlie's measurement result is \ket{0}.

(2) In case (ii), Bob selects to MEAS-RESEND and Charlie selects to REFLECT.
Alice measures particle B in $Z$ basis and particle C in $X$ basis.
After measurements, $\ket{\phi^-}_{BC}$ or $\ket{\psi^+}_{BC}$ collapses to
$\ket{0-}_{BC}$, $\ket{1-}_{BC}$, $\ket{0+}_{BC}$, or $\ket{1+}_{BC}$, each with probability 1/4,
as $\ket{\phi^-}=\frac{1}{\sqrt{2}}(\ket{+-}+\ket{-+})_{BC}$ and $\ket{\psi^+}=\frac{1}{\sqrt{2}}(\ket{++}-\ket{--})_{BC}$.
For example, suppose \ket{\Psi} collapses to $\ket{\phi^-}_{BC}$ after Bob*'s eavesdropping
and $\ket{\phi^-}_{BC}$ changes to $\ket{0-}_{BC}$ or $\ket{1-}_{BC}$ after Alice's measurements.
Alice will found that $\ket{0-}_{BC}$ is not the right state. Thus Bob*'s
eavesdropping will introduce an error rate of 1/2.

(3) In case (iii), suppose \ket{\Psi} collapses to $\ket{\phi^-}_{BC}$ after Bob*'s eavesdropping.
Alice obtains $\ket{+0}_{BC}$, $\ket{+1}_{BC}$, $\ket{-0}_{BC}$, or $\ket{-1}_{BC}$ after her measurements and
$\ket{+1}_{BC}$, $\ket{-0}_{BC}$ are not the right state. In this case, the error rate introduced
by Bob* is also 1/2.

(4) In case (iv), \ket{\Psi} collapses to $\ket{\phi^-}_{BC}$ or $\ket{\psi^+}_{BC}$ after Bob*'s eavesdropping.
Alice performs Hadamard transformation on particle B and the state of particle B and C becomes
\begin{eqnarray}
\frac{1}{\sqrt{2}}(\ket{+0}-\ket{-1})_{BC}=\frac{1}{\sqrt{2}}(\ket{\phi^+}-\ket{\psi^-}),
\end{eqnarray}
or
\begin{eqnarray}
\frac{1}{\sqrt{2}}(\ket{+1}+\ket{-0})_{BC}=\frac{1}{\sqrt{2}}(\ket{\phi^+}+\ket{\psi^-}).
\end{eqnarray}
Alice will found Bob*'s eavesdropping with probability 1/2.

Thus the average error rate introduced by Bob* in this strategy
reaches $\frac{1}{4}*(0+\frac{1}{2}+\frac{1}{2}+\frac{1}{2})=\frac{3}{8}=37.5\%$.

\subsection{Bob*'s entanglement attack}
Suppose Bob* intercepts particle C at step (1) and uses it and his own ancillary particle
in the state $\ket{0}$ to do a CNOT operation (particle C is the controller, Bob*'s ancillary particle is the target).
Bob* then resends particle C to Charlie. Thus
the state of particle B, C and Bob*'s ancillary particle becomes
\begin{eqnarray}
\ket{\Psi_1}&=&\frac{1}{\sqrt{2}}(\ket{+00}+\ket{-11})_{BCB'}=\frac{1}{2}(\ket{000}+\ket{100}+\ket{011}-\ket{111})_{BCB'},
\end{eqnarray}
where $B'$ denotes Bob*'s ancillary particle.
According to the protocol, Bob* and Charlie choose randomly either to MEAS-RESEND or to REFLECT.

(1) In case (i), all of the communication parties perform $Z$ basis measurement and the state collapses to
$\ket{011}_{BCB'}$, $\ket{111}_{BCB'}$, $\ket{000}_{BCB'}$, or $\ket{100}_{BCB'}$, each with probability 1/4.
Bob* will not introduce any error and he can achieve Charlie's secret message.

(2) In case (ii), Bob* chooses to MEAS-RESEND and Charlie chooses to REFLECT. Thus
\ket{\Psi_1} collapses to $\ket{0}_B\ket{\phi^+}_{CB'}$ or $\ket{1}_B\ket{\phi^-}_{CB'}$.
Alice measures particle B in $Z$ basis and particle C in $X$ basis.
In view of $\ket{0}_B\ket{\phi^+}_{CB'}=\frac{1}{\sqrt{2}}\ket{0}_B(\ket{++}+\ket{--})_{CB'}$
and $\ket{1}_B\ket{\phi^-}_{CB'}=\frac{1}{\sqrt{2}}\ket{1}_B(\ket{+-}+\ket{-+})_{CB'}$,
the state of three particles changes to $\ket{0++}_{BCB'}$, $\ket{0--}_{BCB'}$,
$\ket{1+-}_{BCB'}$, or $\ket{1-+}_{BCB'}$ after Alice's measurements.
If Alice obtains $\ket{0-}_{BC}$ or $\ket{1+}_{BC}$,
she detects that there must exist eavesdropping in the transmission line.
Thus the error rate introduced by Bob* is 50\%.

(3) In case (iii), Bob* selects to REFLECT and Charlie selects to MEAS-RESEND.
\ket{\Psi} then collapses to $\ket{+00}_{BCB'}$ or $\ket{-11}_{BCB'}$.
Alice measures particle B in $X$ basis and particle C in $Z$ basis and obtains $\ket{+0}_{BC}$ or $\ket{-1}_{BC}$.
Thus Bob* will not introduce any error and his eavesdropping will not be detected by Alice.

(4) In case (iv), both Bob* and Charlie selects to REFLECT. After Alice's Hadamard transformation,
\ket{\Psi_1} becomes $\ket{\Psi_2}=\frac{1}{\sqrt{2}}(\ket{000}+\ket{111})_{BCB'}$.
As
\begin{eqnarray}
\ket{\Psi_2}&=&\frac{1}{2}[(\ket{\phi^+}+\ket{\phi^-})\ket{0}+(\ket{\phi^+}-\ket{\phi^-})\ket{1}]_{BCB'},
\end{eqnarray}
Alice performs Bell basis measurement on particle B and C, and she will detect the existence of eavesdropping
with probability 50\%.

Thus, in this attack, the average error rate introduced by Bob* is
$\frac{1}{4}*(0+\frac{1}{2}+0+\frac{1}{2})=\frac{1}{4}=25\%$.

At worst, suppose there is an extremely strong eavesdropper who can acquire the communication parties' operation information
by judging whether particle B and C are entangled or not. The communication parties can defeat this attack strategy by utilizing
order rearrangement. It only needs to make some slight modifications to step (2)-(3) of the previous protocol. 
The modified steps are as follows:

(2') When each particle arrives, Bob (Charlie) first selects randomly either to measure it in the computational basis 
and resend it in the same state he found (MEAS-RESEND) or to reflect it to Alice directly (REFLECT). 
Bob (Charlie) then reorders randomly the $B$ ($C$) sequence particles and
generates a rearranged particle sequence. After order rearrangement, Bob (Charlie) sends the rearranged particle sequence to Alice.
The order of the rearranged particle sequence is completely secret to others but Bob (Charlie) himself. 

(3') Alice stores the received particles in quantum memory and informs Bob and Charlie that she has received
the $B$ and $C$ sequence particles. Bob and Charlie then publish which particles he chose to MEAS-RESEND,  
which ones they chose to REFLECT and the secret rearranged order of the particle sequence.

\section{Conclusion}
So far we have proposed a semiquantum secret sharing protocol by using two-particle entangled state
and analyzed the security for the present protocol.
In the protocol, quantum Alice can share a secret key with classical Bob and Charlie. Bob and Charlie
are restricted to measuring a particle in the classical basis,
preparing a particle in the classical basis and sending it, or reflecting a particle directly.
Neither Bob nor Charlie can reconstruct Alice's secret key unless they collaborate. We show the protocol is
secure against eavesdropping because neither Bob nor Charlie can distinguish which particles are MEAS-RESEND particles
and which are REFLECT ones. Even if Bob or Charlie is dishonest, none of them can escape from the eavesdropping check
after acquiring the secret message of the other side. 
Compared with LCL10 protocol, our protocol is simpler because it can be realized by only using two-particle entangled states
instead of three-particle entangled states. As the users only need to perform some classical operations on particles,
semiquantum secret sharing can be realized at a lower cost. We only analyzed the security for the protocol informally
and there are many difficulties to be dealt with when implementing semiquantum secret sharing
in the practical scenario. We would like to explore these problems in the future.

\section*{Acknowledgements}
This work is supported by the National Natural Science Foundation of
China under Grant No. 60872052.



\bibliographystyle{model1a-num-names}
\bibliography{mybib}

\end{document}